\documentclass[twocolumn,final]{svjour3}         
\smartqed  
\usepackage{graphicx}
\usepackage{wasysym}

 \journalname{Nanoscale Research Letters}
\begin{document}

\title{Phase transition on the Si(001) clean surface prepared in UHV MBE chamber\thanks{The research was supported by the Science and Innovations Agency of RF under the State Contract No.~02.513.11.3130 and the Education Agency of RF under the State Contract No.~$\rm \Pi$2367.}
}
\subtitle{A study by high resolution STM and in situ RHEED}

\titlerunning{Phase transition on the Si(001) clean surface prepared in UHV MBE chamber}        

\author{L.~V.~Arapkina   \and
        V.~A.~Yuryev \and
       K.~V.~Chizh \and 
V.~M.~Shevlyuga \and 
M.~S.~Storojevyh \and 
L.~A.~Krylova
}


\institute{Larisa~V.~Arapkina \at
              A. M. Prokhorov General Physics Institute of RAS, 38 Vavilov Street, Moscow, 119991, Russia \\
              Tel.: +7-499-5038318\\
              \email{arapkina@kapella.gpi.ru}           
           \and
           Vladimir~A.~Yuryev \at
              Tel.: +7-499-5038144\\
              Fax: +7-499-1350356\\
              \email{vyuryev@kapella.gpi.ru} 
}

\date{Received: date / Accepted: date}

\maketitle

\begin{abstract}
The Si(001) surface deoxidized by short annealing at  $T \sim 925^{\circ}$C in the ultrahigh vacuum molecular beam epitaxy chamber  has been  {\it in situ} investigated by high resolution scanning tunnelling microscopy  (STM) and reflected high energy electron diffraction (RHEED).   RHEED patterns corresponding to $(2\times 1)$ and $(4\times 4)$ structures were observed during sample treatment.  The $(4\times 4)$ reconstruction arose at $T \apprle 600^{\circ}$C  after annealing. The reconstruction was observed to be reversible: the $(4\times 4)$ structure turned into the $(2\times 1)$ one at $T \apprge 600^{\circ}$C, the $(4\times 4)$ structure appeared again at recurring cooling. The $c(8\times 8)$ reconstruction was revealed by  STM  at room temperature on the same samples. A fraction of the surface area covered by the $c(8\times 8)$  structure decreased as the sample cooling rate was reduced. The $(2\times 1)$ structure was observed on the surface free of the  $c(8\times 8)$ one. The $c(8\times 8)$ structure has been evidenced to manifest itself as the $(4\times 4)$ one in the RHEED patterns. 
A model of the $c(8\times 8)$ structure formation has been built on the basis of the STM data. Origin of the high-order structure on the Si(001) surface and its connection with the epinucleation phenomenon are discussed.
\end{abstract}


\section{\label{sec:intro}Introduction}


Investigations of clean silicon surfaces prepared in conditions of actual technological chambers are of great interest due to the industrial 
requirements to operate
 on nanometer and subnanometer scale when designing 
future nanoelectronic devices \cite{Report_01-303}. In the nearest future, the  sizes of structural elements of such devices will be close to the dimensions of structure features of Si(001) surface, at  least of its high-order reconstructions such as $c(8\times 8)$. Most of  researches of the Si(001) surface have thus far been carried out in specially refined conditions which allowed one to study the most common types of the surface 
reconstructions such as $(2\times 1)$, $c(4\times 4)$,  $c(4\times 2)$  or $c(8\times 8)$ \cite{Hamers,Chadi,Hu,Murray,Kubo,Iton,Koo,Hata,Liu1,Okada,Goryachko,Liu2,our_Si(001)_en}.  
Unfortunately, no or very few papers have thus far been devoted to investigations of the Si surface which is formed as a result of the wafer cleaning and deoxidation directly in the device manufacturing equipment \cite{our_Si(001)_en}.  But anyone who deals with Si-based nanostructure engineering and the development of  such nanostructure formation cycles compatible with some standard device manufacturing processes meets the challenging problem of obtaining the clean Si surface within the imposed technological restrictions which  is one of the key elements of the entire structure formation cycle 
\cite{Report_01-303,photon-2008ag,Si(001)_ICDS-25}.

The case is that the ambient in technological vessels such as molecular beam epitaxy (MBE) chambers  is usually not as pure as in specially refined  ones designed for surface studies. There are many sources of surface contaminants in the process chambers including materials of wafer heaters or evaporators of elements as well as foreign substances used for epitaxy and doping. 
In addition, due to technological reasons the temperature treatments applicable for device fabrication following the standard processes such as CMOS often cannot be as aggressive as those used for surface preparation in the basic experiments. Moreover, the commercially available technological equipment sometimes does not enable the wishful annealing of Si wafers at the temperature of $\sim 1200\,^\circ$C  even if the early device formation stage allows one to heat the wafer to such a high temperature. 
Nevertheless, the technologist should always  be convinced that the entirely deoxidized and atomically clean Si surface is reliably and reproducibly obtained. 

A detailed knowledge of the Si surface structure which is formed in the above conditions---its reconstruction, defectiveness, fine structural peculiarities, etc.---is of great importance too because this structure may affect the properties of nanostructured layers deposited on it.  For instance, the Si surface structure may affect the magnitude and the distribution of the surface stress   of the Ge wetting layer on nanometer scale when the Ge/Si structure is grown, which in turn affect the Ge nanocluster nucleation and eventually the properties of quantum dot arrays formed on the surface \cite{Report_01-303,Si(001)_ICDS-25,Nanophysics,Ion_Dvur,Ion_irradiation_1,Dvur_PLDS,Dvur_irrad,Dvur_mod,Dvur_ion-beam-epi,Dvur_book_ion,Ion_irradiation,Smagina,classification,atomic_structure,Hut_nucleation,defects_ICDS-25}. 

Thus, it is evident from the above that the controllable formation of the clean Si(001) surface with the prescribed parameters required for technological cycles of nanofabrication compatible with the standard device manufacturing processes should be considered as an important goal, and this article presents a step to it.

In the present paper, we report the results of investigation of the Si(001) surface treated following the standard procedure of Si wafer preparation for the MBE growth of the SiGe/Si(001) or Ge/Si(001) heterostructures. A  structure  arising on the Si(001) surface as a result of short high-temperature annealing for SiO$_2$ removal
is explored.  It is well known that such
experimental treatments favor the formation of nonequilibrium structures on the surface. The most studied of them are presently the $(2\times 1)$ and $c(4\times 4)$ ones. 
This work experimentally investigates by means of scanning tunneling microscopy (STM) and reflected high energy electron diffraction (RHEED)
the formation and atomic structure of the less studied high-order  $c(8\times 8)$ (or  $c(8\times n)$ \cite{our_Si(001)_en,photon-2008ag,Si(001)_ICDS-25}) reconstruction. Observations of  this reconstruction have already been reported in the literature \cite{Hu,Murray,Kubo,Liu1} but there is no  clear comprehension of causes of its formation as the structures looking like the $c(8\times 8)$ one appear after different treatments: The  $c(8\times 8)$ reconstruction was observed to be a result of the coper atoms deposition on the Si(001)-$(2\times 1)$ surface \cite{Iton,Liu1}; similar structures were found to arise due to various treatments and low-temperature annealing of the original Si(001)-$(2\times 1)$ surface without deposition of any foreign atoms  \cite{Hu,Murray,Kubo}. Data of the STM studies of the Si(001)-$c(8\times 8)$ surface were presented in Refs.~\cite{Murray,Liu1}.

It may be supposed on the analogy with the Si(001)-$c(4 \times 4)$ reconstruction  \cite{Goryachko,Miki,Urberg,c(4x4),surface_morphology,oxygen_induced_ordered}   that the presence of impurity atoms on the surface as well as in the subsurface regions is not the only reason of formation of reconstructions different from the $(2\times 1)$ one, and the conditions of thermal treatments should be taken into account. The results of exploration of effect of such factor as the rate of sample cooling  from the annealing temperature to the room one
on the process of the $c(8\times 8)$ reconstruction formation 
are reported  in the present article. It is shown by means of RHEED that the diffraction patterns corresponding to the $(2\times 1)$ surface structure reversibly turn into those corresponding to the $c(8\times 8)$ one depending on the sample temperature, and a point of this phase transition is determined. Based on the STM data a model of the $c(8\times 8)$ structure formation is brought forward.

\section{Methods and equipment}
\label{sec:setup}
The experiments were made using an integrated ultra-high-vacuum (UHV) system \cite{classification} based on the Riber EVA\,32 molecular beam epitaxy chamber equipped with the Staib Instruments RH20 diffractometer of reflected high energy electrons and coupled through a transfer line with the GPI~300 UHV     scanning tunnelling microscope \cite{gpi300,STM_GPI-Proc,STM_calibration}. This instrument enables the STM study of samples at any stage of Si surface preparation and MBE growth. The samples can be serially moved into the STM chamber for the analysis and back into the MBE vessel for further treatments  as many times as required never leaving the UHV ambient. RHEED experiments can be carried out {\it in situ}, i.e. directly in the MBE chamber during the process. 

Samples for STM were 8$\times$8 mm$^{2}$ squares cut from the specially treated commercial B-doped    CZ Si$(100)$ wafers ($p$-type,  $\rho\,= 12~\Omega\,$cm). RHEED measurements were carried out at the STM samples and similar $2''$ wafers; the $2''$ samples were investigated only by means of RHEED.  
After chemical treatment following the standard procedure described elsewhere \cite{Report_01-303,etching} (which included washing in ethanol, etching in the mixture of HNO$_3$ and HF and rinsing in the deionized water), the samples were placed in the holders.
The STM samples were  mounted on the molybdenum STM holders and inflexibly clamped with the tantalum fasteners. The STM holders were placed in the  holders for MBE made of molybdenum with tantalum inserts.  
The $2''$ wafers were inserted directly into the standard molybdenum  MBE holders and did not have so hard  fastening as the STM samples.

\begin{figure}
\includegraphics[scale=2]{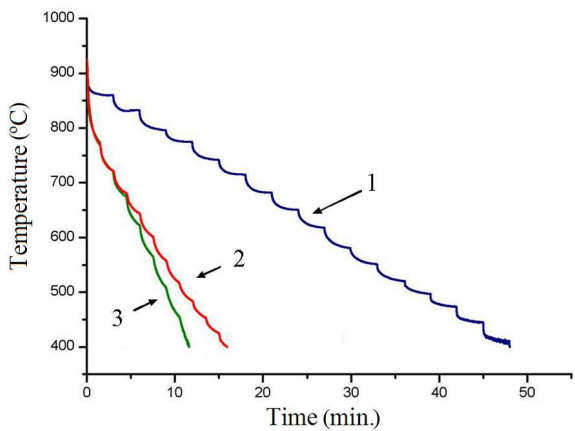}
\caption{\label{fig:cooling_rate} A diagram of sample cooling after the thermal treatment at $925^{\circ}$C measured by IR pyrometer; cooling rates are as follows: $\sim 0.17^{\circ}$C/s or ``slow cooling'' of the STM samples (1); $\sim 0.4^{\circ}$C/s or ``quenching" of the STM samples (2) and $2''$ wafers (3).
}
\end{figure}

 Thereupon the samples were loaded into the  airlock and transferred into the preliminary annealing chamber where outgassed at  $\sim 600\,^\circ$C and  $\sim 5\times 10^{-9}$ Torr for about 6 hours. After that the samples were moved  for final treatment and decomposition of the oxide film into the MBE chamber evacuated down to $\sim 10^{-11}$\,Torr. There were two stages of annealing in the process of sample heating\,---at $\sim 600\,^\circ$C for $\sim 5$\,min and at $\sim 800\,^\circ$C for $\sim 3$\,min \cite{Report_01-303,our_Si(001)_en,classification}.  The final annealing was carried out at  $\sim 925\,^\circ$C.\footnote{ The samples were heated over $920\,^\circ$C about a half of the final annealing time; a diagram of the thermal processing and some additional details can be found in Ref.\,\cite{classification}.} Then the temperature was rapidly lowered to $\sim 850\,^\circ$C. The rates of the further cooling down to the room temperature were  $\sim 0.4\,^\circ$C/s (referred to as the ``quenching'' mode of  both the STM samples and $2''$ wafers) or  $\sim 0.17\,^\circ$C/s (called the ``slow cooling'' mode of only the STM samples) (Fig.~\ref{fig:cooling_rate}). The pressure in the MBE chamber increased  to $\sim 2\times 10^{-9}$ Torr during the process.

 In both chambers, the samples were heated from the rear side by  radiators of tantalum. The temperature was monitored with the IMPAC~IS\,12-Si pyrometer which measured the Si sample temperature through chamber windows.
The atmosphere composition in the MBE chamber was monitored using the SRS~RGA-200 residual gas analyser before and during the process.

After cooling, the STM samples were moved into the STM chamber in which the pressure did not exceed  $1\times 10^{-10}$ Torr. 
RHEED patterns were obtained for all samples directly in the MBE chamber at different elevated temperatures in the process of the sample treatment and at room temperature after cooling. The STM samples were always explored by RHEED before moving into the STM chamber.

The STM tips were {\it ex situ} made of the tungsten wire and cleaned by ion bombardment \cite{W-tip}  
in a special UHV chamber connected to the STM chamber. The STM images were obtained in the constant tunnelling current 
mode at room temperature. The STM tip was zero-biased while the sample was positively or negatively biased 
when scanned in empty or filled states imaging mode. 


The STM images were processed afterwords using the WSxM software \cite{WSxM}.

\section{\label{sec:results}Experimental findings}

\begin{figure}
\includegraphics[scale=1.23]{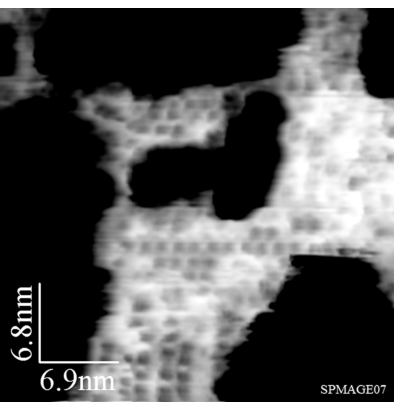}(a)
\includegraphics[scale=0.7]{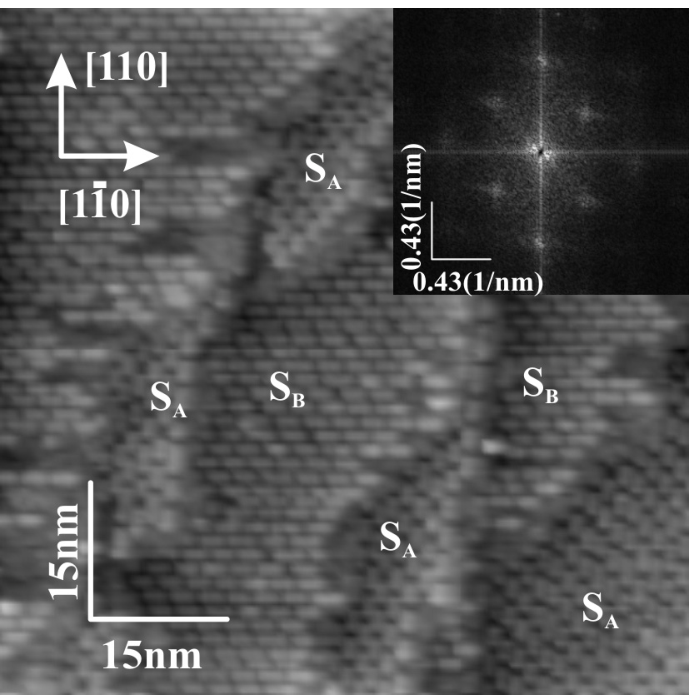}(b)
\caption{\label{fig:stm}
STM image of the Si(001) surface with the residual silicon oxide ($-1.5$~V, 150~pA), annealing at $\sim 925^\circ$C for $\sim 2$~min. (a), the image is inverted: dark areas correspond to the oxide, the lighter areas represent the deoxidized surface; STM image of the clean Si(001) surface ($+1.9$~V, 70~pA) with the Fourier transform pattern shown in the insert, annealing at $\sim 925^\circ$C for $\sim 3$~min. (b) \cite{our_Si(001)_en}.
}
\end{figure}

\begin{figure*}
5
\includegraphics[scale=1.2]{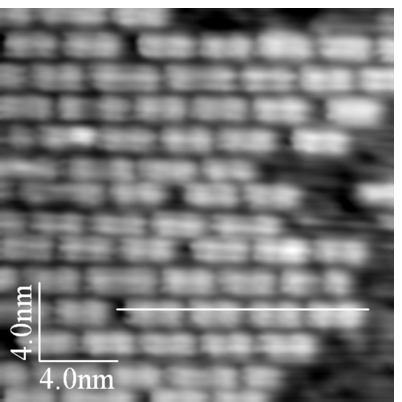}(a)
\includegraphics[scale=1.2]{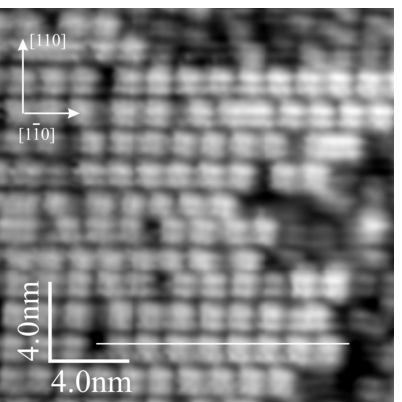}(b)
\includegraphics[scale=1.2]{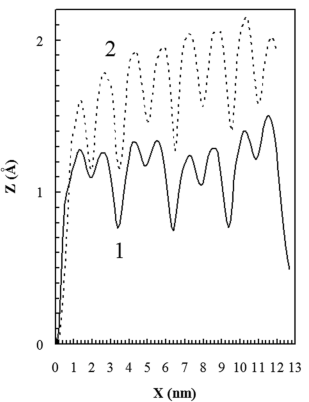}(c)
\caption{\label{fig:profile}
Empty (a) and filled (b) state  images of the same region on the Si(001) surface (+1.7~V, 100~pA and $-2.0$~V, 100~pA). Positions of extremes  of line scan profiles (c) match exactly for the empty (1) and filled (2) state distributions along the corresponding lines in the images (a) and (b).
}
\end{figure*}

Fig.~\ref{fig:stm} demonstrates STM images of the Si(001) surface after annealing at $\sim 925^{\circ}$C of different duration.  Fig.~\ref{fig:stm}a depicts the early phase of the oxide film removal; the annealing duration is 2 min. A part of the surface is still oxidized: the dark areas in the image correspond to the surface coated with the oxide film. The lighter areas correspond to the purified surface. A structure of ordered ``rectangles'' (the grey features) is observed on the deoxidizes surface. After longer annealing (for 3 min.) and quenching (Fig.~\ref{fig:cooling_rate}), the surface is entirely purified of the oxide (Fig.~\ref{fig:stm}b). 
It consists of terraces separated by the $\mathrm{S_B}$  or $\mathrm{S_A}$ monoatomic steps  with the height of $\sim 1.4$~\r{A} \cite{Chadi}.  Each terrace is composed of rows running along $[110]$ or $[1\overline{1}0]$ directions. The surface reconstruction is different from the $(2\times 1)$ one. The insert of Fig.~\ref{fig:stm}b demonstrates the Fourier transform of this image which corresponds to the $c(8\times 8)$ structure \cite{Murray}: Reflexes of the Fourier transform 
correspond to the distance $\sim 31$~\r{A}  in both $[110]$ and $[1\overline{1}0]$ directions. So the revealed structure   have a periodicity of $\sim 31$~\r{A} that corresponds to 8 translations $a$ on the surface lattice of Si(001) along the $<$110$>$ directions ($a=3.83$~\r{A} is a unit translation length).  Rows consisting of structurally arranged rectangular  blocks are clearly seen in the empty state STM image (Fig.~\ref{fig:stm}b).
They turn by $90^{\circ}$  on the neighbouring terraces.

Fig.~\ref{fig:profile} demonstrates the empty  and filled  state images of the same surface region. Each block consists of two lines separated by a gap. This fine structure of blocks is clearly seen in the both pictures (a) and (b) but its images are different in different scanning modes. A characteristic property most clearly seen  in the filled state mode (Fig.~\ref{fig:profile}b) is the presence of the brightness maxima on both sides of the lines inside the blocks. These peculiar features are described below in more detail. 
Fig.~\ref{fig:profile}c shows the profiles of the images taken along the white lines. Extreme positions of both curves are  well fitted. Relative heights of the features outside and inside the blocks can be estimated from the profiles.

\begin{figure}
\includegraphics*[scale=0.9]{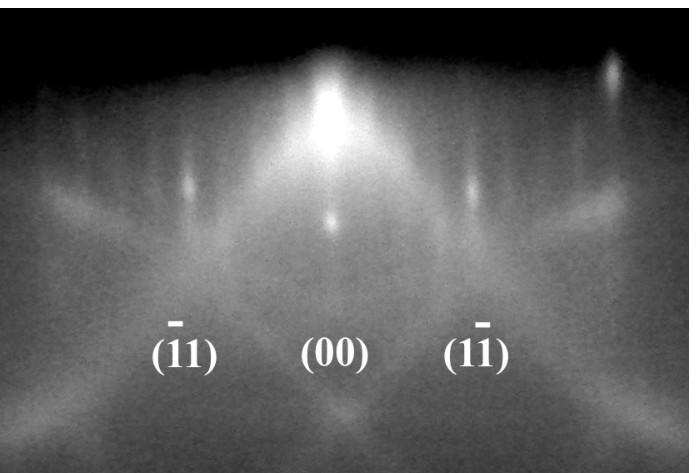}(a) 
\includegraphics*[scale=0.9]{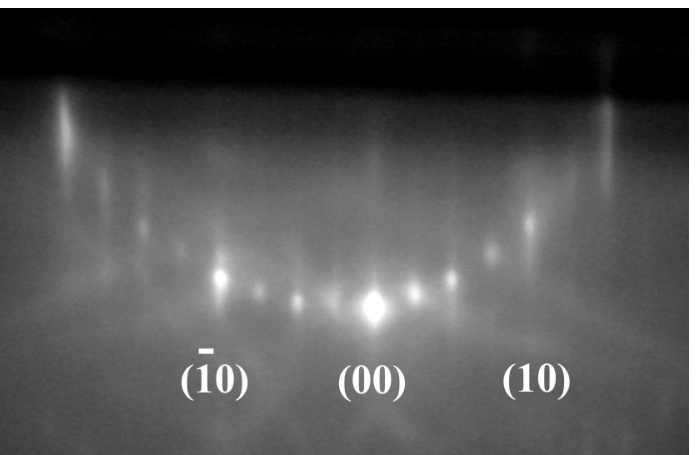}(b)
\caption{\label{fig:rheed}
Reflected high energy electron diffraction patterns observed in the $[0\,1\,0]$  (a) and $[1\,1\,0]$  (b) azimuths; electron energy was 9.8~keV and 9.3~keV, respectively.
}
\end{figure}  

Fig.~\ref{fig:rheed} demonstrates typical RHEED patterns taken at room temperature from the STM sample annealed for 3 min. with further quenching. Characteristic  distances on the surface corresponding to the reflex positions in the diffraction pattern  were calculated according to Ref.~\cite{geometrical_fundamentals}. The derived surface structure is $(4\times 4)$. 
One sample showed the RHEED patterns corresponding to the $(2\times 1)$ structure \cite{geometrical_fundamentals} after the same treatment though.

Temperature dependences of the RHEED patterns in the $[$110$]$ azimuth  were investigated during sample heating and cooling. It was found that the reflexes corresponding to $2a$ were distinctly seen  in the RHEED patterns  during annealing at  $\sim 925^{\circ}$C after 2 minutes of treatment. The reflexes corresponding to $4a$ started to appear during sample quenching and became definitely visible at the temperature of $\sim 600^{\circ}$C; a weak $(4\times 4)$ signal started to arise at $\sim 525^{\circ}$C if the sample was cooled slowly (Fig.~\ref{fig:cooling_rate}). At the repeated heating from room temperature to $925^{\circ}$C, the $(4\times 4)$ structure disappeared at   $\sim 600^{\circ}$C giving place to the $(2\times 1)$ one. The $(4\times 4)$ structure  appeared again at $\sim 600^{\circ}$ during recurring cooling.  

The RHEED patterns obtained  from $2''$ samples always corresponded to the $(2\times 1)$ reconstruction. Diffraction patterns for the STM sample which was not hard fastened to the holder corresponded to the $(4\times 4)$ structure after quenching (STM measurements were not made for this sample).

\begin{figure}
\includegraphics[scale=0.9]{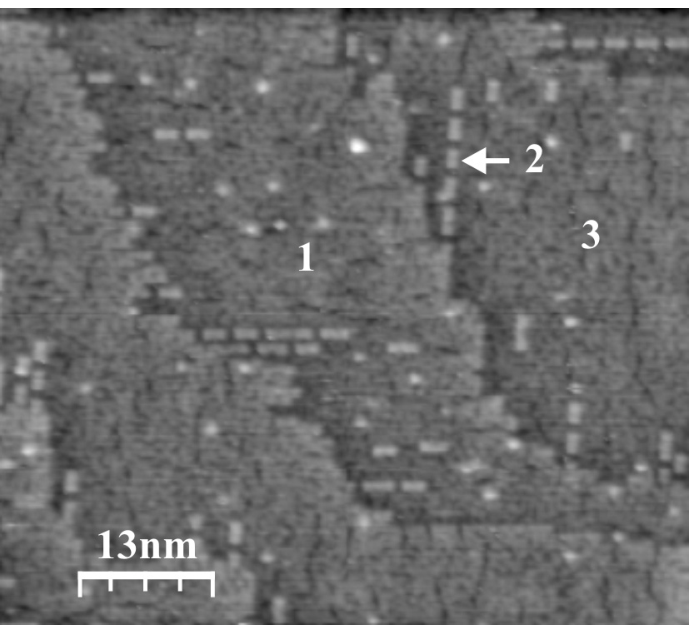}(a)
\includegraphics[scale=0.9]{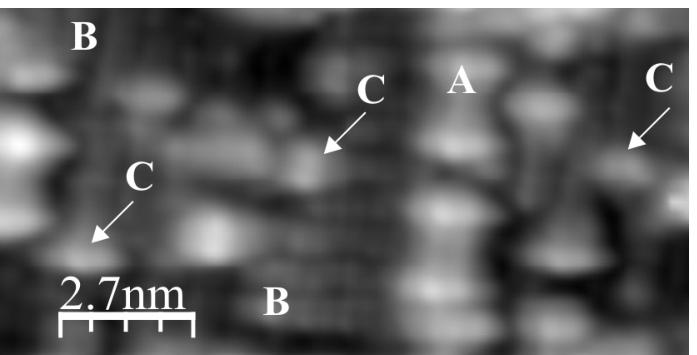}(b)
\caption{\label{fig:slow_cooling}
STM images of the clean Si(001) surface prepared in the slow cooling mode: (a) the surface mainly covered by the $(2\times 1)$ structure, +2.0~V, 100~pA, `1' and `3' are terraces, the height of the row `2' coincides with the height of the terrace `1'; 
  a magnified image taken with atomic resolution (b), 
$-1.5$~V, 150~pA, `A' is the ``rectangle'', `B' marks the dimer rows composing the $(2\times 1)$ structure ({\it separate atoms are seen}), `C' shows structural defects, i.e. 
the dimers of the uppermost layer oriented along the dimers of the lower $(2\times 1)$ rows (b).
}
\end{figure}

Effects of annealing duration and cooling rate 
on the clean surface structure were studied by STM. It was established that increase of annealing duration to 6 min. did not cause any changes of the surface structure. On the contrary,  decrease of the sample cooling rate drastically changes the structure of the surface. The STM images of the sample surface for the slow cooling mode (Fig.~\ref{fig:cooling_rate}) are presented in
Fig.~\ref{fig:slow_cooling}. The difference of this surface from that of the quenched samples  (Fig.~\ref{fig:stm}b) is that only a few rows of ``rectangles'' are observed on it.  The order of the ``rectangle'' positions with the period of $8a$ remains in such rows. Two adjacent terraces are designated in Fig.~\ref{fig:slow_cooling}a by figures `1' and `3'. A row of ``rectangles'' marked as `2' is situated on the terrace `3'; it has the same height as the terrace `1'. The filled state image, which is magnified in comparison with the former one, is  given in Fig.~\ref{fig:slow_cooling}b. A part of the surface free of the ``rectangles'' is occupied by the $(2\times 1)$ reconstruction. Images of the dimer rows with the resolved  Si atoms  are marked as `B' in Fig.~\ref{fig:slow_cooling}b. The ``rectangles'' are also seen in the image (they are marked as `A') as well as single defects: dimerized Si atoms (`C') and chaotically located on the surface accumulations of several dimers. Most of these dimers are oriented parallel to dimers of the lower surface and located strictly on the dimer row. Note that influence of the cooling rate on the surface structure was observed by the authors of Ref.~\cite{Kubo}: when the sample cooling rate was decreased the surface reconstruction turned from $c(8\times 8)$ to $c(4\times 2)$ which was considered as the derivative reconstruction  of the $(2\times 1)$ one transformed because of  dimer buckling. 

\begin{figure}
\includegraphics[scale=0.9]{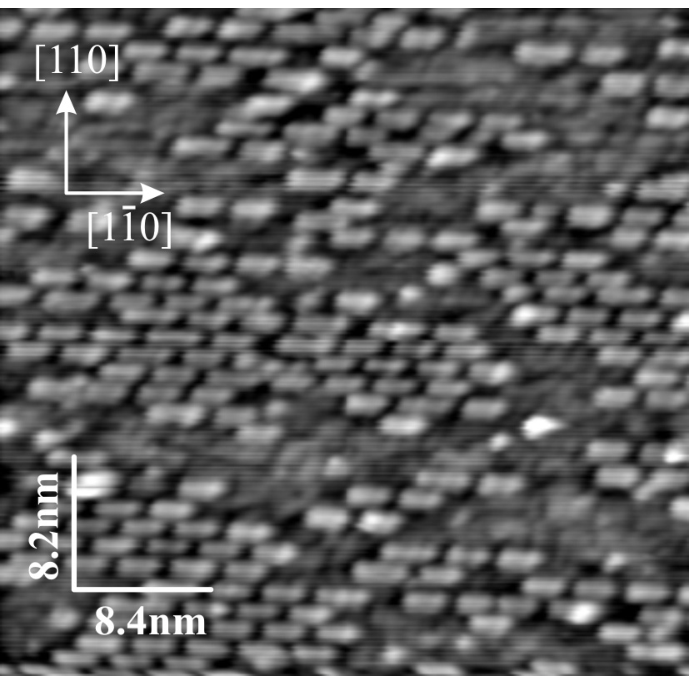}(a)
\includegraphics[scale=0.90]{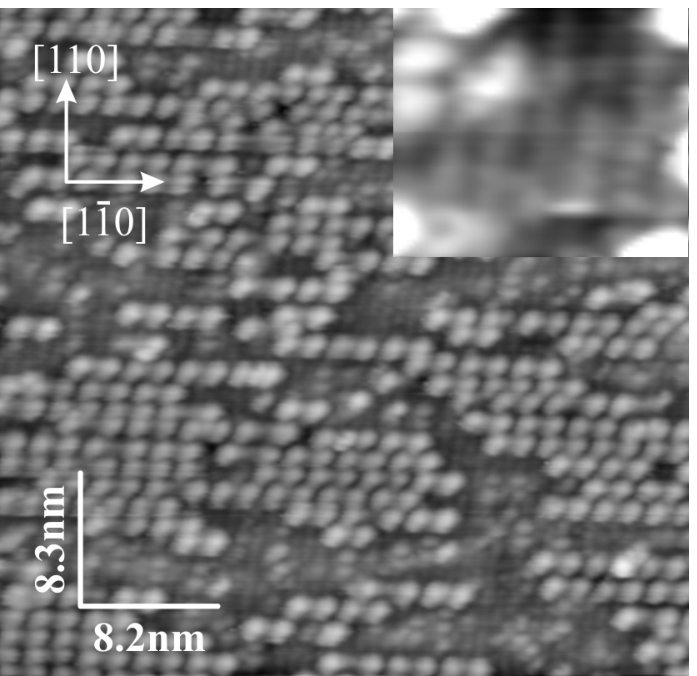}(b)
\caption{\label{fig:empty-filled} 
Empty (a) and filled (b) state  images of the same region on the Si(001) surface  (+2.0~V, 100~pA and $-2.0$~V, 100~pA); an insert at (b) shows the image of the $(2\times 1)$ surface obtained between the rows of ``rectangles''.
}
\end{figure}

Fig.~\ref{fig:empty-filled} presents the STM images obtained for the samples cooled in the quenching mode  but containing areas free of ``rectangles''. The images (a) and (b) of the same place on the surface were obtained serially one by one. We managed to image the surface structure between the areas occupied by the ``rectangle'' rows, but only in the filled state mode (see the insert at Fig.~\ref{fig:empty-filled}b).  Like in Fig.~\ref{fig:slow_cooling}b this structure is seen to be formed by parallel dimer rows going $2a$ apart. The direction of these rows is perpendicular to the direction of the rows of ``rectangles''. The height difference of the rows of ``rectangles'' and the $(2\times 1)$ rows is 1 monoatomic step ($\sim 1.4$\,\r{A}).
We did not succeed to obtain a good enough image of these subjacent dimer rows in the empty state mode. It should be noted also that positions of the ``rectangles'' are always  strictly fixed 
relative to the dimer rows of the lower layer: they  occupy exactly three subjacent dimer rows. It also may be seen in the STM images presented in Refs.~\cite{Murray,Liu1}.

\subsection{\label{sec:details}Fine structure of the observed reconstruction }

Let us consider the observed structure  in detail.

A purified sample surface consists of monoatomic steps. Following the nomenclature by Chadi \cite{Chadi}, they are designated as $S_{\rm A}$ and $S_{\rm B}$ in Fig.~\ref{fig:stm}b. Each terrace is composed by rows running along the $[110]$ or $[1\overline{1}0]$ directions.  Each row consists of rectangular blocks (``rectangles''). They may be regarded as surface structural units as they are present on the surface after thermal treatment in any mode,  irrespective of a degree of surface coverage by them.
Reflexes of the Fourier transform of the picture shown in  Fig.~\ref{fig:stm}b correspond to the distances $\sim 31$ and $\sim 15$\,\r{A} in both [110] and $[1\overline{1}0]$   directions. Hence the structure revealed in the long shot seems to have a periodicity of $\sim 31$\,\r{A} that corresponds to 8 translations $a$ on the surface lattice of Si(001). It resembles the Si(001)-$c(8\times 8)$ surface  \cite{Murray}. Reflexes corresponding to the distance of $\sim 15$\,\r{A} ($4a$) arise due to the periodicity along the rows. 
STM images obtained at higher magnifications give an evidence that the surface appears to be disordered, though.

\begin{figure}
\includegraphics[scale=0.8]{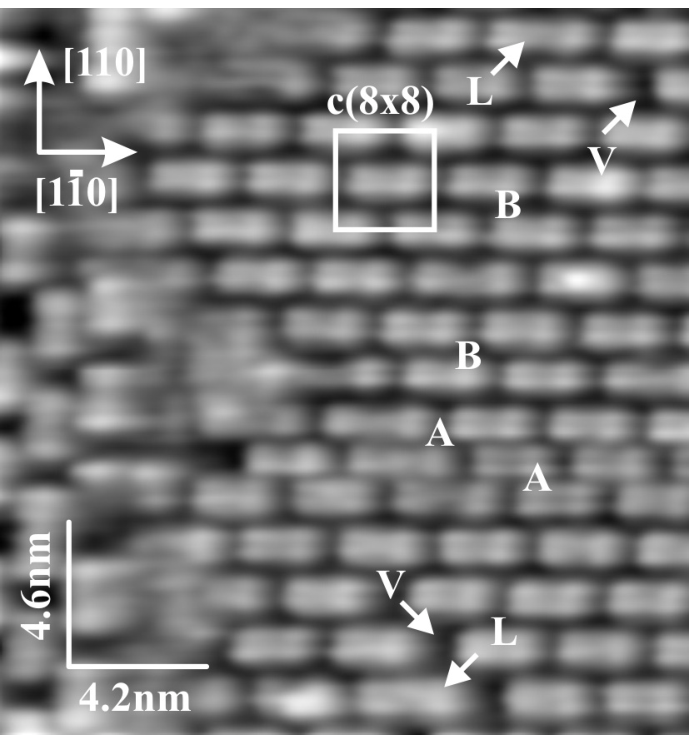}(a)
\includegraphics[scale=1.4]{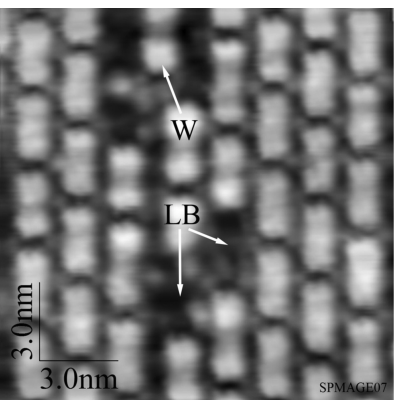}(b)
\caption{\label{fig:8x8} 
STM empty state images of the  Si(001) surface ;  a $c(8\times 8)$ unit cell is marked by the white box in image (a) ($+1.9$~V, 50~pA), distances between the rows marked by `A' and `B' equal  $3a$  and $4a$ (that corresponds to $c(8 \times 6$) and $c(8 \times 8$) structures, respectively), two long ``rectangles'' and divacancies arising in the adjacent rows are marked by `L' and `V', respectively; a row wedging between two rows (`W') and  lost blocks (`LB') are seen in (b) ($+1.6$~V, 100~pA).  
 }
\end{figure}

Fig.~\ref{fig:8x8} shows the magnified images of the investigated surface. The rows of the   blocks are seen to be situated at varying distances from one another (hereinafter, the distances are measured between corresponding maxima of features). A unit $c(8\times 8)$ cell is marked with a square box in Fig.~\ref{fig:8x8}a. The distances between the adjacent rows of the  rectangles  are $4a$
in such structures (`B' in Fig.~\ref{fig:8x8}a).  The adjacent rows designated as `A' are $3a$ apart ($c(8\times 6)$).  

A structure with the rows going at $4a$ apart is presented in Fig.~\ref{fig:8x8}b. The lost blocks (`LB') that resemble point defects are observed in this image. In addition, a row wedging in between two rows and separating them by an additional distance $a$   is seen in the centre of the upper side of the picture (`W'). The total distance between the wedged off rows becomes $5a$.

Hence it may be concluded that the order and some periodicity take place only along the rows---disordering of the $c(8\times 8)$ structure across the rows is revealed (we often refer to this structure as $c(8\times n)$).

\begin{figure}
\includegraphics[scale=0.75]{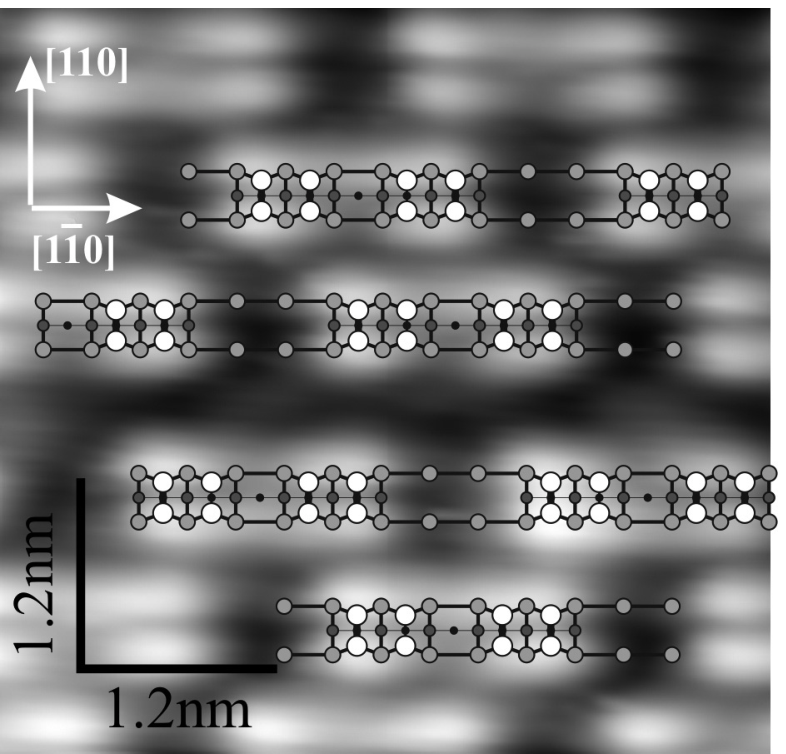}(a)
\includegraphics[scale=0.75]{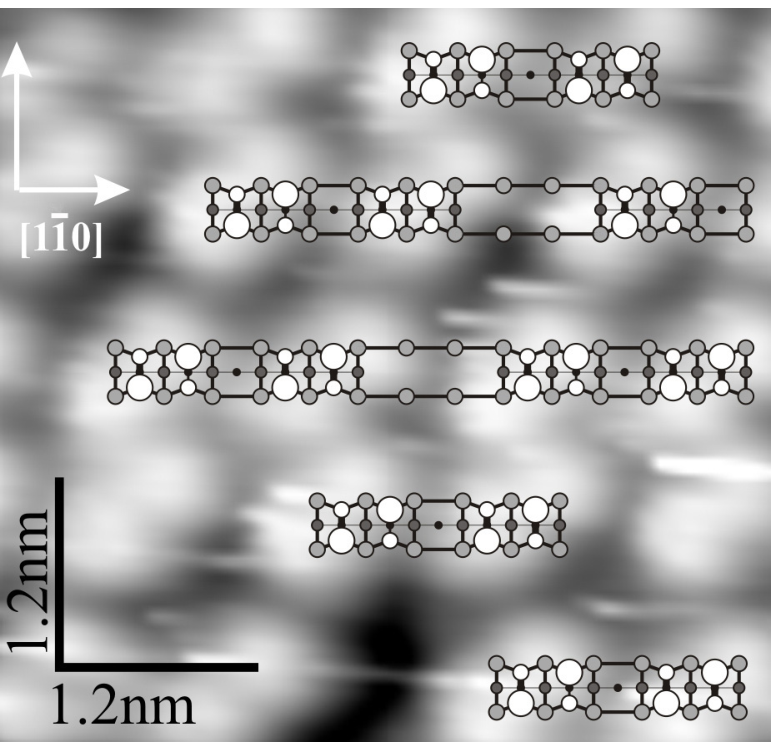}(b)
\caption{\label{fig:block} 
Empty state  (a) and filled state  (b) images of the  Si(001)-$c(8\times n)$ surface ($+1.7$~V, 150~pA, and $-2.2$~V, 120~pA). Corresponding schematic drawings of the surface structure are superimposed on both pictures. The lighter circle is the higher the corresponding atom is situated in the surface structure. The dimer buckling is observed in the filled state image, which is reflected in the drawing by larger open circles representing  higher  atoms of the tilted Si dimers of the uppermost layer of the structure.
}
\end{figure}

The block length can possess two values: $\sim 15$\,\r{A} ($4a$) and $\sim 23$\,\r{A} ($6a$). 
Distances between equivalent positions of the adjacent short blocks in the rows are $8a$. If the long block appears in a row, a divacancy is formed in the adjacent row to restore the checkerboard order of blocks. Fig.~\ref{fig:8x8}a illustrates this peculiarity. The long block is marked as `L', the divacancy arisen in the adjacent row is lettered by `V'.
In addition, the long blocks were found to have one more peculiarity. They have extra maxima in their central regions. The maxima are not so pronounced as the main ones but nevertheless they are quite recognizable in the pictures (Fig.~\ref{fig:8x8}a).

Fig.~\ref{fig:block} presents  magnified STM images of the blocks (``short rectangles''). The images obtained in  the empty-state (Fig.~\ref{fig:block}a) and filled-state (Fig.~\ref{fig:block}b) modes are different. In the empty-state mode, short blocks look like two  lines separated by $\sim 8$\,\r{A} (the distance is measured between  brightness maxima in each line). It is a maximum measured value which can lessen depending on scanning parameters. Along the rows, each block is formed by two parts. The distance between the brightness  maxima of these parts is $\sim 11.5$\,\r{A} (or some greater depending on scanning parameters). In the filled-state mode, the block division into two structurally identical parts remains. Depending on scanning conditions, each part looks like either bright coupled dashes and blobs (Figs.~\ref{fig:profile}b and~\ref{fig:empty-filled}b) or two links (brightness maxima) of zigzag chains (Fig.~\ref{fig:block}b). The distances between the maxima  are $\sim 4$\,\r{A} along the rows. 

The presented STM data are interpreted by us as a structure composed by Si ad-dimers and divacancies.

\section{\label{sec:discussion}Discussion}

\subsection{\label{sec:model}Structural model }

The above data  allow us to bring forward a model of the observed Si(001) surface reconstruction. The model is based on the following assumptions: (i) the outermost surface layer is formed by ad-dimers; (ii) the underlying layer has a structure of $(2\times 1)$; (iii) every rectangular block consists of ad-dimers and divacancies a number of which controls the block length.

\begin{figure*}
\includegraphics*[scale=1.4]{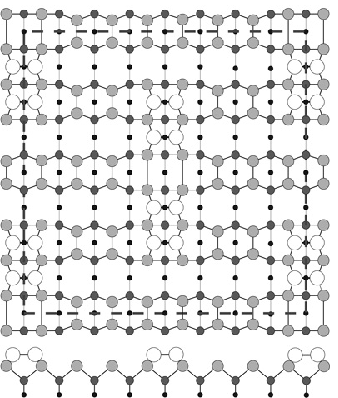}(a)
\includegraphics*[scale=1.4]{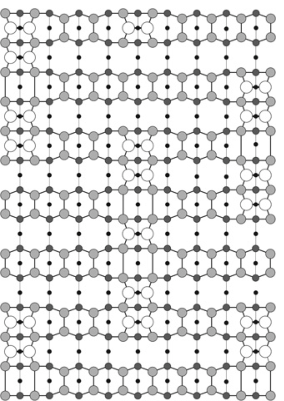}(b)
\includegraphics*[scale=1.4]{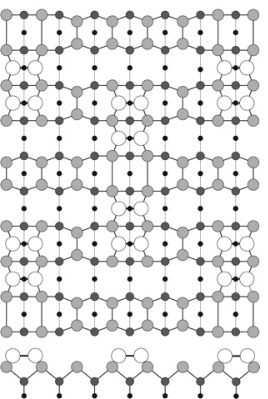}(c)
\caption{\label{fig:8xn}
A schematic drawing of the $c(8\times n)$ structure:  $c(8\times 8)$ with the short blocks (a), a unit cell is outlined; the same structure with the long block (b);  $c(8\times 6)$ structure  (c).
}
\end{figure*}

Fig.~\ref{fig:8xn}a shows a schematic drawing of the $c(8\times 8)$ structure (a unit cell is outlined). This structure is a basic one for the model brought forward. The elementary structural unit is a short rectangle. These blocks form raised rows running vertically (shown by empty circles). Smaller shaded circles show horizontal dimer rows of the lower terrace. The rest black circles show bulk atoms. Each ``rectangle'' consists of two  dimer pairs separated with a dimer vacancy. The structures on the Si(001) surface composed of close ad-dimers are believed to be stable \cite{Kubo,Liu2} or at least metastable \cite{epinucleation}. In our model, a position of the ``rectangles'' is governed by the location of the dimer rows of the  $(2\times 1)$ structure of the underlying layer. The rows of blocks are always normal to the dimer rows in the underlying layer to form a correct epiorientation \cite{epinucleation}. Every rectangular block is bounded by the dimer rows of the underlying layer from both short sides. Short sides of blocks form non-rebonded $S_{\rm B}$ steps \cite{Chadi} with the underlying substrate (see  Fig.~\ref{fig:slow_cooling}b and three vertically running (the very left) rows of ``rectangles'' in  Fig.~\ref{fig:8x8}a).

Fig.~\ref{fig:8xn}b demonstrates the same model for the case of the long rectangle. This block is formed due to the presence of an additional dimer in the middle of the rectangle. The structure consisting of one dimer is metastable \cite{Kubo,Liu2}, so this type of blocks cannot be dominating in the structure. Each long block is bounded on both short sides by the dimer rows of the underlying terrace, too. The presence of the long rectangle results in the formation of a dimer-vacancy defect in the adjacent row; this is shown in Fig.~\ref{fig:8xn}b---the long block is drawn in the middle row, the dimer vacancy is present in the last left row.
According to our STM data the surface is disordered in the direction perpendicular to the rows of the blocks. The distances between the neighboring rows may be less than those in the $c(8\times 8)$ structure. Hence the structure presented in this paper may be classified as $c(8\times n)$ one. Fig.~\ref{fig:8xn}c demonstrates an example of such  structure---the $c(8\times 6)$ one. 

In Fig.~\ref{fig:block}, the  presented structure is superimposed on  STM images of the surface. The filled state image (Fig.~\ref{fig:block}b) reveals dimer buckling in the blocks which is often observed in this  mode at some values of sample bias and tunnelling current. Upper atoms of tilted dimers are shown by  larger open circles.

\begin{figure}
\includegraphics*[scale=1.3]{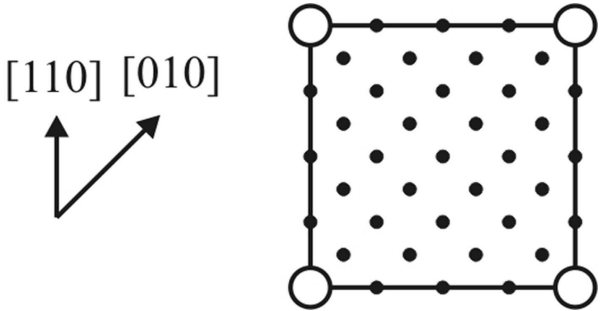}
\caption{\label{fig:lattice} 
The  Si(001)-$c(8 \times 8)$ surface reciprocal lattice.
}
\end{figure}

\begin{figure*}
\includegraphics[scale=0.75]{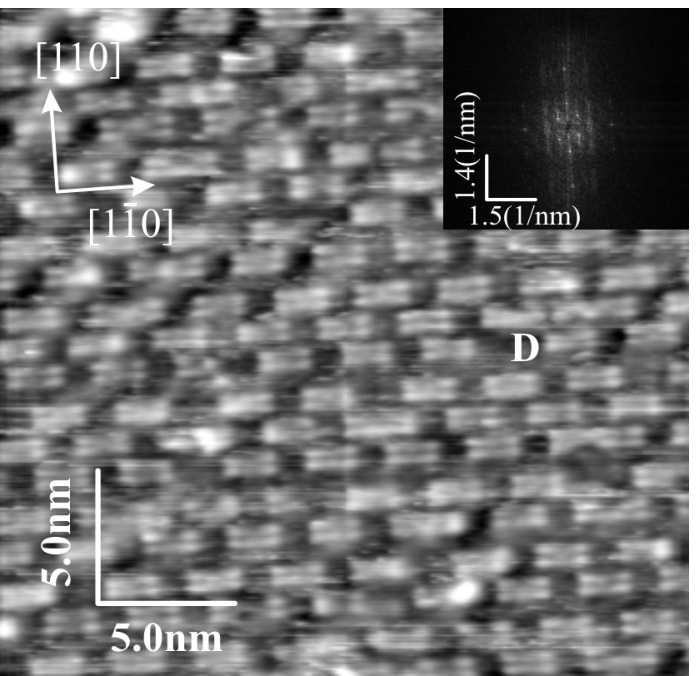}(a)
\includegraphics[scale=0.75]{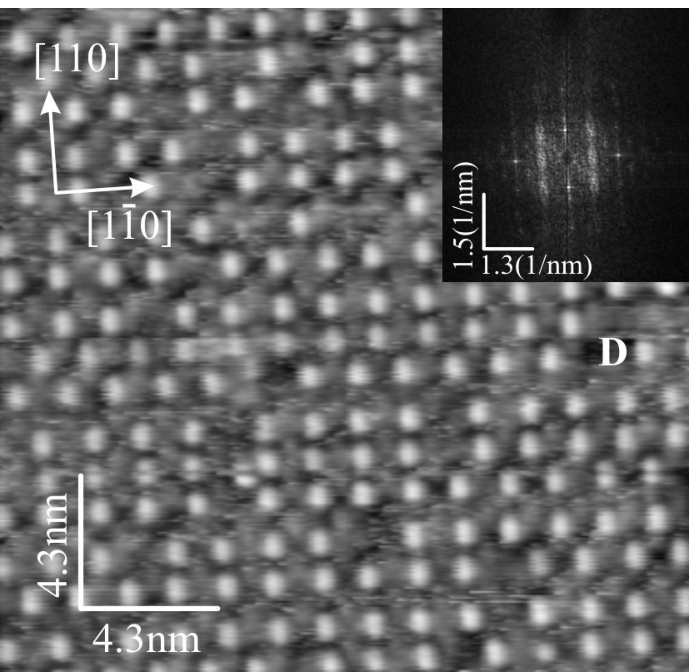}(b)
\includegraphics[scale=1.1]{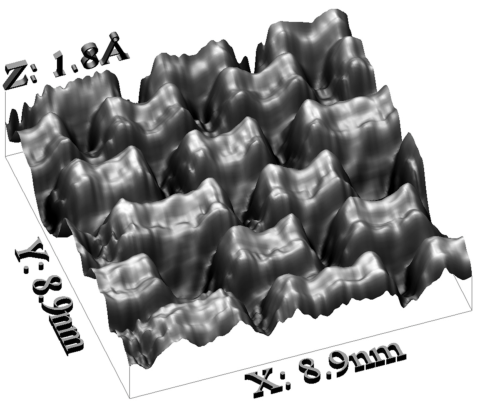}(c)
\caption{\label{fig:4x4} 
STM images of the same area on the  surface obtained in the empty state (a) and frilled state  (b) modes ($+1.96$~V, 120~pA and $-1.96$~V, 100~pA); for the convenience of comparison, `D' indicates the same vacancy defect; corresponding Fourier transforms are shown in the inserts. A 3-D STM empty state micrograph (+2.0~V, 200~pA) of the Si(001)-$c(8\times 8)$ surface is shown in (c).
}
\end{figure*}

\subsection{\label{sec:comparison}Comparison of STM and RHEED data}

Now d discrepancy of results obtained by STM and RHEED  within the proposed model. 
Fig.~\ref{fig:lattice} presents a sketch of the reciprocal lattice of $c(8\times 8)$. The RHEED patterns obtained in the $[110]$ azimuth correspond to the  $c(8\times 8)$ structure; the patterns observed in the $[010]$ azimuth do not (Fig.~\ref{fig:rheed}). The reason of this discrepancy may be understood from the STM filled state image which corresponds to the electron density distribution of electrons paired   in 
covalent bond of a Si--Si dimer. Fig.~\ref{fig:4x4} compares STM images of the same region on one terrace obtained in the empty-state (a)  and filled-state (b) modes;  inserts show their Fourier transforms, the differences in which for the two STM modes are as follows: in the Fourier transform of the filled state image, reflexes corresponding to the distance of $8a$ are absent in the $[110]$ and $[1\overline{1}0]$ directions, whereas the reflexes corresponding to $4a$ and $2a$ are present (it should be noticed that the image itself resembles that of the $(4\times 4)$ reconstructed surface). If an empty state image is not available, it might be concluded that the $(4\times 4)$ structure is arranged on the surface. An explanation of this observation is simple. Main contribution to the STM image is made by ad-dimers situated on the sides of the ``rectangles'', i.\,e. on tops of the underlying dimer rows. According to calculations made, e.\,g., in  Refs.~\cite{energetics_dynamics,energetics_adatoms} 
dimers located in such a way are closer to the STM tip and look in the images brighter than those situated in the troughs. Hence, it may be concluded that the RHEED $(4\times 4)$  pattern  results from electron diffraction on the extreme dimers of the ``rectangles'' forming the $c(8\times 8)$ surface structure.

The latter statement is illustrated by the  STM 3-D empty-state topograph shown in Fig.~\ref{fig:4x4}c. The extreme dimers located on the sides of the rectangular blocks are seen to be somewhat higher than the other ones of the dimer pairs; they form a superfine relief which turned out to be sufficient to backscatter fast electrons incident on the surface at grazing angles.

\subsection[Origin]{Origin}
\label{sec:origin}

The Si(001)-$c(8\times 8)$ structure have formerly been observed and described in a number of publications \cite{Hu,Murray,Kubo,Iton,Liu1}. Conditions of its formation were different: we shall  explain the observe coper atoms were deposited on silicon $(2\times 1)$ surface to form the $c(8\times 8)$ reconstruction \cite{Liu1}, although it is known that Cu atoms are not absorbed on the Si(001) clean surface if the sample temperature is greater than $600^{\circ}$C, and on the contrary Cu desorption from the surface takes place ~\cite{Iton,Liu1}; fast  cooling  from the annealing temperature of $\sim 1100^{\circ}$C  was applied \cite{Hu,Murray};  samples treated in advance by ion bombardment were annealed and rapidly cooled \cite{Kubo}. The resultant surfaces were mainly explored by STM and low energy electron diffraction (LEED). STM investigations yielded alike results---a basic unit of the reconstruction was a ``rectangle'', but the  structure of the ``rectangles'' revealed by different authors was different. In general, an origin of the Si(001)-$c(8\times 8)$ structure  is unclear now.

STM images most resembling our data were reported in Ref.~\cite{Murray}. 
In that paper, the $c(8\times 8)$ structure was observed in  samples without special treatment by coper: the samples were subjected to annealing at the temperature of $\sim 1050^\circ$C for the oxide film removal. Formation of the  $c(8\times 8)$ reconstruction was explained in that article by the presence of a trace amount of Cu atoms the concentration of which was beyond the Auger electron spectroscopy detection threshold. The STM empty state  images of the samples were similar to those presented in the current paper. A very important difference is observed in the filled state images---we observe absolutely different configuration of dimers within the ``rectangles''.
Nevertheless, the presence of Cu cannot be completely excluded. Some amount of the Cu atoms may come on the surface from the construction materials of the MBE chamber (although there is a circumstance that to some extent contradicts this viewpoint:  Cu atoms were not detected in the residual atmosphere of the MBE chamber within the sensitivity limit of the SRS RGA-200 mass spectrometer) or even from the Si wafer. Cu is known to be a poorly controllable impurity and its concentration in the subsurface layers of Si wafers which were not subjected to the gettering process may reach ~10$^{15}$~cm$^{-3}$. 
This amount of Cu may appear to be sufficient to give rise to the formation of the defect surface reconstruction. However, the following arguments urge us to  doubt about the Cu-based model: (i) undetectable trace amounts of Cu were suggested  in Ref.~\cite{Murray}, the presence or absence of which is unprovable; (ii) even if the suggestion is true, our STM images give an evidence of a different amount of dimers in the rectangular blocks, so, it is unclear why Cu atoms form different stable configurations on similar surfaces; and (iii) it is hard to explain why Cu atoms cyclically compose and decompose the rectangular blocks during the cyclical thermal treatments of the samples. It applies equally to any other impurity or contamination. 

Now we consider a different interpretation of our data. As mentioned above,
 literature suggests two causes of  $c(8\times 8)$ appearance. The first is an impact of impurity atoms adsorbed on the surface even at trace concentrations. The second is a thermal cycle of the oxide film decomposition and sample cooling. The first model seems to be hardly applicable for explanation of the reported experimental results. 
According to our  data, there are no impurities adsorbed directly on the studied surface: RHEED patterns correspond to a clean Si(001) surface reconstructed in $(2\times 1)$ or, at lower temperatures, $(4\times 4)$ configuration. Cyclic contaminant desorption at high temperatures ($\apprge 600^{\circ}$C) and adsorption on sample cooling is unbelievable. Consecutive segregation and desegregation of an undetectable impurity in subsurface layers also does not seem verisimilar.

\begin{figure*}
\includegraphics*[scale=1.5]{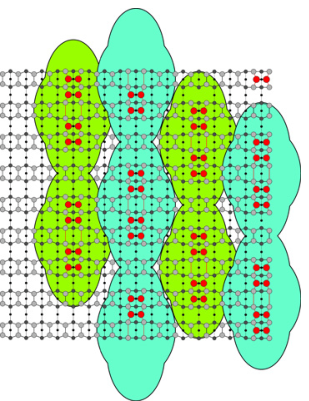}(a)
\includegraphics*[scale=1.5]{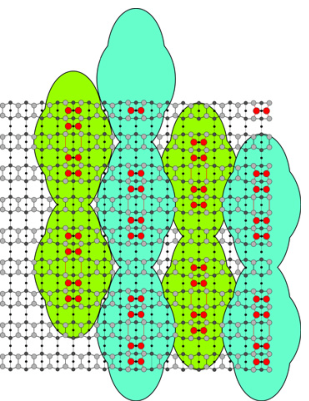}(b)
\includegraphics*[scale=1.5]{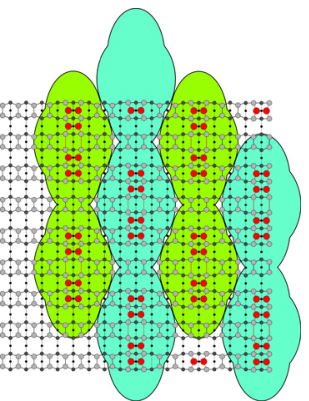}(c)
\caption{\label{fig:strain} 
Schematic representation of the surface stress fields interactions during formation of the $c(8\times 8)$ structure: (a) ordering of the ``rectangles'' within the rows; (b) ordering of the rows relative to each other; (c) the ordered $c(8\times 8)$ structure.
}
\end{figure*}

The second explanation looks more attractive.
It was found in Ref.~\cite{Ney} as a result of the STM studies that the Si(001) surface subjected to the thermal treatment at  $\sim 820^\circ$C which was used for decomposition of the thin ($\sim 1$~nm thick) SiO$_2$ films obtained by chemical oxidation contained a high density of vacancy-type defects and their agglomerates as well as individual ad-dimers. So, the initial  bricks for the considered surface structure are abundant after the SiO$_2$ decay.

Literature presents a wide experimental material on a different reconstruction of the Si(001) surface---$c(4\times 4)$---which also arise at the temperatures of $\apprge 600^{\circ}$C. For example, a review of articles describing different experimental investigations can be found in Refs.~\cite{Goryachko,Miki,Urberg,c(4x4),surface_morphology,oxygen_induced_ordered}. Based on the generalized data, an inference can be made that the   $c(4\times 4)$ structure forms in the interval from 600 to 700$^{\circ}$C. Most likely, at these temperatures an appreciable migration of Si ad-atoms  starts on surface. The structure is free of impurities.
It irreversibly transits to the $(2\times 1)$ one at the temperature greater than $720^{\circ}$C. 
Ref.~\cite{ordered-defects} demonstrates formation of the Si(001)-$(2\times 8)$ structure, also without impurity atoms. In analogy with the above literature data,  formation of the $c(8\times 8)$ reconstruction may be expected as a result of low-temperature annealing and/or further quenching. The standard annealing temperature for obtaining $(2\times 1)$ structure is known to be in the interval from 1200 to 1250$^{\circ}$C. At these temperatures in UHV ambient not only oxide film removal from the surface takes place, but also silicon evaporation  and carbon desorption goes on.  Unfortunately, we have not got a technical opportunity 
to carry out such a high-temperature annealing in our instrument. Treatment at 925$^{\circ}$C that we apply likely does not result in   substantial evaporation of Si atoms from the surface, and C atoms, if any, may diffuse into subsurface layers. As a result, a great amount of ad-dimers arise on the surface, like it happens in the process described in Ref.~\cite{Ney}. Formation processes of the $(2\times 1)$ and $c(8\times 8)$ structures are different.  $(2\times 1)$ arise during the high-temperature annealing and ad-atoms of the uppermost layer do not need to migrate and be embedded into the lattice to form this reconstruction. On the contrary, $c(8\times 8)$ appears during sample cooling, at rather low temperatures, and at the moment of a prior annealing the uppermost layer consists of abundant ad-atoms. On cooling the ad-dimers have to migrate along the surface and be build in the lattice. A number of competing sinks may exist on the surface (steps, vacancies, etc.), but high cooling rate may impede ad-atom annihilation slowing their migration to sinks and in such way creating supersaturation and favoring 2-D islanding, and freezing a high-order reconstruction.

The following scenario may be proposed to describe the $c(8\times 8)$ structure formation. A large number of ad-dimers remains on the surface during the sample annealing after the oxide film removal. They form the uppermost layer of the structure. The underlying layer is $(2\times 1)$ reconstructed. Ad-dimers are mobile and can form different complexes (islands). Calculations show that the most energetically favorable island configurations are single dimer on a row in non-epitaxial orientation \cite{epinucleation,energetics_adatoms,self-organization_Si,atomic_molecular} (Fig.~\ref{fig:slow_cooling}b), complexes of two dimers (pairs of dimers) in epi-orientation (metastable \cite{epinucleation}) and two dimers on a row in non-epitaxial orientation separated by a divacancy, and tripple-dimer epi-islands considered as critical epinuclei \cite{epinucleation}. These mobile dimers and complexes migrate in the stress field of the  $(2\times 1)$ structure. The sinks for ad-dimers are (A) steps, (B) vacancy defects of the underlying  $(2\times 1)$ reconstructed layer, and (C) ``fastening'' them to   the $(2\times 1)$ surface as a $c(8\times 8)$ structure. The main sinks at high temperatures are A and B. As the sample is cooled, the C sink becomes dominating. Ad-dimers on the  Si(001)-$(2\times 1)$ surface are known to tend to form dimer rows \cite{growth_equilibrium}. In this case such rows are formed by metastable dimer pairs gathered in the ``rectangles''. The ``rectangles'' are ordered with a period of 8 translations in the rows. The ordering is likely controlled by the $(2\times 1)$ structure of the underlying layer and interaction of the stress fields arising around each ``rectangle''. Effect of the underlying $(2\times 1)$ layer is that the ``rectangle'' position on the surface  relative to its dimer rows is strictly defined:   dimers of the ``rectangle'' edges must be placed on tops of the rows. Interaction of the stress fields initially arranges the ``rectangles'' within the rows (Fig.~\ref{fig:strain}a), then it arranges adjacent rows with respect to one another (Fig.~\ref{fig:strain}b). The resultant ordered structure is shown in Fig.~\ref{fig:strain}c. The described behaviour of ``rectangles'' can be derived from the STM images presented in the previous sections. In addition, investigation of appearance of the RHEED patterns allowed us to conclude that the process of dimer ordering in the $c(8\times 8)$ structure is gradual: the pattern reflexes appearing on transition from $(2\times 1)$ to $(4\times 4)$ reach maximum brightness gradually; it means that the  $c(8\times 8)$ structure does not arise instantly throughout the sample surface, but originally form some nuclei (``standalone rectangles'' like those in Fig.~\ref{fig:slow_cooling}a) on which mobile  ad-dimers crystallize in the ordered surface configuration.

\subsection[Stability]{Stability}
\label{sec:stability}

A source of stability of the Si(001) surface configuration composed by ad-dimers gathered in the rectangular islands has not been found to date. Some of possible sources of stabilization of structures with high-order periodicity  were considered in Refs.~\cite{Miki,ordered-defects,dimer_vacancy_array,missing_dimers,c(4x8)}. One of likely reasons of high-order structure formation might be a non-uniformity of the stress field distribution on a sample surface and dependance of this distribution on such factors as process temperature, sample cooling rate, specimen geometry and a way of sample  
fastening to a holder, presence of impurity atoms on and under the surface. In this wise, it is clear only that ad-dimers form ``rectangles'' which are energetically favorable at temperature conditions of the experiments. 

In this connection, a guide for further consideration could be found in Ref.~\cite{epinucleation} where an issue of the critical epinucleus---or the smallest island which unreconstructs the surface and whose probability of growth is greater than likelihood of decay---on the $(2\times 1)$ reconstructed Si(001) was theoretically investigated. First-principle calculations showed that dimer pairs in epi-orientation are metastable and the epinucleus consists of tripple dimers \cite{epinucleation}. 
Unfortunately, we failed to observe tripple-dimer islands in our experiments, and calculations were limited to three dimers in the cited article. Some formations smaller than ``rectangles'' sometimes are observed  in images of the rarified structures (Fig.~\ref{fig:slow_cooling}a) but they are likely single dimers (Fig.~\ref{fig:slow_cooling}b) and dimer pairs. We believe that
the short ``rectangles'' we deal with in this article  might be considered as epinuclei for the $c(8\times n)$ structure because, although they show no tendency to grow themselves, they are both seeds and structural units for formation of larger islands such as chains  (Fig.~\ref{fig:slow_cooling}a), grouped chains (Fig.~\ref{fig:stm}a) and complete ares (Fig.~\ref{fig:empty-filled}). From other hand, they also do not tend to decay or  annihilate even on as powerful sinks as steps (Fig.~\ref{fig:slow_cooling}a). Thus, we conclude that the stability of such epi-islands as dimer pair-vacancy-pair (short ``rectangles'', Fig.~\ref{fig:8xn}a,c) is the highest. Less probable (stable)  configuration is pair-vacancy-dimer-vacancy-pair (long ``rectangle'', Fig.~\ref{fig:8xn}b). We think its less stability is due to presence of a single epi-oriented dimer in the centre. That is why long ``rectangles'' are  much less spread on the Si(001) surface than the short ones and entire structure stabilization in the presence of the long ``rectangles'' requires appearance of additional dimer vacancies between ``rectangles'' in adjacent rows in the vicinity of the long blocks.

\subsection{Remark on connection with Ge epitaxial growth}
\label{sec:Ge-epi}
We would like to notice  that the temperature interval from 550 to 600$^{\circ}$C, in which the reported phase transition occurs, is commonly adopted as a frontier between the so-called low-temperature and high-temperature modes of Ge quantum dot array growth on the Si(001) surface \cite{Dvur_UFN}. This means that  low-temperature arrays obtained by MBE usually grow on the $c(8\times n)$  reconstructed Si surface densely covered by the described above ``rectangles'' if no special precautions are taken to ensure slow cooling of a Si substrate after surface preparation for Ge  deposition. High-temperature arrays always form on the   $(2\times 1)$ reconstructed surface. The difference in the initial surface morphology may cause a difference in stress distribution in Ge wetting layer which, in turn, may affect the cluster nucleation and  growth. Of course, this hypothesis requires an accurate experimental checkup.

\section{Conclusion}
\label{sec:conclusion}

In summary, it may be concluded that the Si(001) surface prepared under the conditions of the UHV MBE chamber in a standard wafer preparation cycle has $c(8\times n)$ reconstruction which is partly ordered only in one direction. Two types of unit blocks form the rows running along $[110]$ and $[1\overline{1}0]$ axes. When the long block disturbs the order in a row a dimer-vacancy defect appears in the adjacent row in the vicinity of the long block to restore the checker-board order of blocks in the neighboring rows.

Discrepancy  of RHEED patterns and STM images was detected. According to RHEED data, 
 $(2\times 1)$ and $(4\times 4)$ structures can form the Si(001) surface during sample treatment. STM studies of the same samples at room temperature  show that
a high-order $c(8\times 8)$ reconstruction exists on the Si(001)  surface; simultaneously, the underlying layer is   $(2\times 1)$ reconstructed in the areas free of the $c(8\times 8)$ structure. 
 A fraction of the surface area covered by the $c(8\times 8)$  structure decreases as the sample cooling rate is reduced. 
RHEED patterns corresponding to  the $(4\times 4)$ reconstruction arise at $\sim 600^{\circ}$C in the process of sample cooling after annealing. The reconstruction is  reversible: the $(4\times 4)$ structure turns into the $(2\times 1)$ one at $\sim 600^{\circ}$C in the process of the repeated sample heating, the $(4\times 4)$ structure appears on the surface again at the same temperature during recurring cooling. 

A model of the $c(8\times 8)$ structure  based on  epi-oriented ad-dimer complexes
has been presented. Ordering of the ad-dimer complexes likely arises due to interaction of the stress fields produced by them. 
The discrepancy of the STM and RHEED data has been explained within the proposed model: 
the $c(8\times 8)$ structure revealed by STM has been evidenced to manifest itself as the $(4\times 4)$ one in the RHEED patterns.

Probable causes of the $c(8\times 8)$ reconstructed Si(001) surface formation have been discussed. 
A combination of low temperature of sample annealing   and high   rate of  its cooling may be considered as one of the most plausible factors responsible for  its appearance.
The structural units of the studied reconstruction are supposed to be its critical epinuclei.



%
%

\bibliography{stm-rheed-EMRS}

\end{document}